\UseRawInputEncoding

\documentclass{article}
\usepackage{amsfonts}
\usepackage{makeidx}
\usepackage{latexsym,amsmath,amssymb,amscd}
\usepackage{makecell}
\usepackage{xcolor}
\usepackage{multirow}
\usepackage[all]{xy}
\usepackage{float}
\usepackage{longtable}

\allowdisplaybreaks
\newtheorem{theorem}{Theorem}

\begin{document}

\title{Higher order first integrals of autonomous dynamical systems}
\author{Antonios Mitsopoulos$^{1,a)}$ and Michael Tsamparlis$^{1,b)}$  \\
%EndAName
{\ \ }\\
$^{1}${\textit{Faculty of Physics, Department of
Astronomy-Astrophysics-Mechanics,}}\\
{\ \textit{University of Athens, Panepistemiopolis, Athens 157 83, Greece}}
\vspace{12pt} %EndAName
\\
$^{a)}$Author to whom correspondence should be addressed: antmits@phys.uoa.gr %EndAName
\\
$^{b)}$Email: mtsampa@phys.uoa.gr }
\date{}
\maketitle

\begin{abstract}
A theorem is derived which determines higher order first integrals of autonomous holonomic dynamical systems in a general space, provided the collineations and the Killing tensors --up to the order of the first integral-- of the kinetic metric, defined by the kinetic energy of the system, can be computed. The theorem is applied in the case of Newtonian autonomous conservative dynamical systems of two degrees of freedom, where known and new integrable and superintegrable potentials that admit cubic first integrals are determined.
\end{abstract}

\bigskip


\section{Introduction}

\label{introduction}

In general, a system of differential equations is integrable if there exist `enough' in number first integrals (FIs) so that its solution can be found by means of quadratures. It is well-known \cite{Arnold 1989} that in the special case of autonomous Hamiltonian systems with $n$ degrees of freedom the above definition becomes more specific. Indeed, such systems are integrable if they admit $n$ (functionally) independent autonomous FIs $I(q,p)$ which are in involution. The last definition is carried over \cite{Kozlov 1983}, \cite{Vozmishcheva 2005} unchanged for non-autonomous Hamiltonian systems $H(q,p,t)$ and general time-dependent FIs $I(q,p,t)$. This means that time-dependent FIs can be used to establish the integrability of a dynamical system. The maximal number of \emph{autonomous} independent FIs is $2n-1$; however, if additional time-dependent FIs exist, this maximal limit can be surpassed.

If there exist $2n-1$ independent FIs, an integrable Hamiltonian system $H(q,p,t)$ is called (maximally) superintegrable. If there are $k$ independent FIs such that $n < k < 2n -1$, the system is called minimally superintegrable. In the case of non-autonomous Hamiltonian systems, the Hamiltonian is not a FI.

A general first order autonomous system $\dot{x}_{i} =F_{i}(x)$, where $i=1,...,n$ and $F_{i}$ are arbitrary smooth functions of the variables $x_{i}$, is always integrable if there exist $n-1$ independent FIs \cite{Yoshida II}. However, the existence of fewer FIs may also be sufficient since in the case of Hamiltonian systems, where $n=2m$, $m$ independent FIs in involution are enough for (Liouville) integrability.

In the course of time, there have been developed various methods which determine FIs. A brief review of the major such methods has as follows.

\subsubsection*{The Lie symmetry method}

A Lie symmetry of a differential equation is a point transformation in the
solution space of the equation which preserves the set of solutions of the
equation. The vector field\footnote{%
We restrict our considerations to the first jet bundle $J^{1}\{t,q,\dot{q}\}$, i.e. to contact transformations.} $\mathbf{X}= \xi(t,q,\dot{q})\partial _{t} +\eta^{a}(t,q,\dot{q}) \partial_{q^{a}}$ which generates the point transformation is called the generator of the Lie symmetry. If the components $\xi(t,q), \eta^{a}(t,q)$, i.e. they do not depend on the `velocities' $\dot{q}^{a}$, the Lie symmetry
is called a point symmetry; otherwise it is a generalized Lie symmetry. A generalized Lie symmetry has one free parameter which is removed by means of a gauge condition. The standard gauge condition is $\xi =0$. Although a Lie symmetry is possible to lead to a FI (see e.g. \cite{Katzin 1973}), in
general it does not, and one has to restrict to Noether symmetries. These are Lie symmetries which satisfy the additional requirement of the Noether condition. Every Noether symmetry leads to a Noether FI. The method of Noether symmetries is the most widely used tool for the determination of FIs (see e.g. \cite{Katzin 1974}, \cite{Katzin 1976}, \cite{Ranada 1997}, \cite{Damianou 2004}, \cite{Mei 2014}, \cite{Hadler}).

\subsubsection*{The inverse Noether theorem}

If $I(t,q,\dot{q})$ is a FI of a second order dynamical system whose Lagrangian $L(t,q,\dot{q})$ is regular, that is, $\det(\gamma_{ab})\neq0$ where $\gamma_{ab}\equiv \frac{\partial^{2} L}{\partial \dot{q}^{a} \partial \dot{q}^{b}}$ is the kinetic metric, then by means of the inverse Noether theorem one may associate to $I$ a gauged generalized Noether symmetry and finally compute the FIs. This is done as follows. From the inverse Noether theorem, the FI $I$ is associated to the generalized Noether symmetry (see e.g. \cite{Djukic 1975}, \cite{Sarlet 1981}, \cite{Tsamparlis 2020B})
\begin{eqnarray}
\eta^{a} &=& -\gamma^{ab} \frac{\partial I}{\partial \dot{q}^{b}} +\xi\dot{q}^{a}  \label{Cartan.1} \\
\xi &=& \frac{1}{L} \left( f - I + \gamma^{ab} \frac{\partial L}{\partial \dot{q}^{a}} \frac{\partial I}{\partial \dot{q}^{b}} \right) \label{Cartan.1.1}
\end{eqnarray}%
where $f(t,q,\dot{q})$ is the Noether function, the Einstein summation convention is used, $\dot{q}^{a}\equiv \frac{dq^{a}}{dt}$ and the kinetic metric $\gamma_{ab}$ is used for lowering and raising the indices. Equation (\ref{Cartan.1}) is the well-known Cartan condition. In the gauge $\xi =0$ conditions (\ref{Cartan.1}), (\ref{Cartan.1.1}) become
\begin{eqnarray}
\eta^{a} &=& -\gamma^{ab} \frac{\partial I}{\partial \dot{q}^{b}} \label{Cartan.1.2} \\
f &=& I -\gamma^{ab} \frac{\partial L}{\partial \dot{q}^{a}} \frac{\partial I}{\partial \dot{q}^{b}}. \label{Cartan.1.3}
\end{eqnarray}
If one looks for quadratic FIs (QFIs) of the form
\begin{equation}
I=K_{ab}(t,q)\dot{q}^{a}\dot{q}^{b}+K_{a}(t,q)\dot{q}^{a}+K(t,q)
\label{Cartan.2}
\end{equation}%
where $K_{ab}(t,q), K_{a}(t,q), K(t,q)$ are symmetric tensor quantities, then from conditions (\ref{Cartan.1.2}), (\ref{Cartan.1.3}) follows that the generator $\eta_{a}= -2K_{ab}\dot{q}^{b}-K_{a}$ and the Noether function $f= -K_{ab}\dot{q}^{a}\dot{q}^{b} +K$. Replacing these into the Noether condition one obtains a set of partial differential equations (PDEs) whose solution provides the corresponding Noether integrals \cite{Leach 1985}.

\subsubsection*{The Lax pair method}

In this method (see e.g. \cite{Lax 1968}, \cite{Calogero}, \cite{Olshanetsky}, \cite{Babelon 1990}, \cite{ArutyunovB}), one brings the dynamical equations into a special matrix form called a Lax representation. Then the existence of an extended set of FIs is guaranteed. Specifically, Hamilton's equations have to be written in the form
\begin{equation}
\dot{A}= [B,A]= BA -AB \label{lax1}
\end{equation}
where $A, B$ are two square matrices whose entries are functions on the phase space $q, p$ of the system. If this is possible, then it is said that the system admits a Lax representation with $A$ being the corresponding Lax matrix. The pair of matrices $A, B$ is called a Lax pair.

If one finds a Lax representation, then the functions
\begin{equation}
I_{k}= tr(A^{k}) \label{lax2}
\end{equation}
where $tr$ denotes the trace and $k$ is a positive integer, are FIs. Indeed, we have
\[
\dot{I}_{k}= k ~tr\left( A^{k-1} \dot{A} \right) = k ~tr\left( A^{k-1} [B,A] \right)= k ~tr\left( A^{k-1}BA \right) -k ~tr\left( A^{k} B\right) = k ~tr\left( A^{k}B \right) -k ~tr\left( A^{k} B\right) = 0
\]
because the trace is invariant under cyclic permutations and $tr(A+B)= tr(A) + tr(B)$. In fact, the matrix equation (\ref{lax1}) has the general solution
\begin{equation}
A(t)= F(t) A(0) F(t)^{-1} \label{lax3}
\end{equation}
where the invertible matrix $F(t)$ is such that $B= \dot{F} F^{-1}$.

A Hamiltonian system may admit more than one Lax pair. These pairs may be 1) represented by square matrices of different size; 2) related by transformations of the type
\begin{equation}
A' = G A G^{-1}, \enskip B'= GBG^{-1} +\dot{G}G^{-1} \label{lax4}
\end{equation}
where $G$ is an arbitrary invertible matrix.

\subsubsection*{ The Hamilton-Jacobi (H-J) method}

This is also a widely applied method and --as a rule-- concerns autonomous conservative dynamical systems and FIs with small degrees of freedom. In this method, one
considers in the phase space (cotangent bundle) the Hamiltonian $H= \frac{1}{2}\gamma^{ab}(q)p_{a}p_{b} +V(q)$ where $V(q)$ denotes the potential and $q^{a}, p_{a}$ are the canonical coordinates. The coordinates $q^{a}$ and the
Hamiltonian are called separable if the corresponding Hamilton-Jacobi (H-J) equation
\begin{equation*}
\frac{1}{2}\gamma^{ab}W_{,a}W_{,b} +V =h
\end{equation*}%
has a complete solution of the form
\begin{equation*}
W(q;c)=W_{1}(q^{1};c) + ... +W_{n}(q^{n};c)
\end{equation*}%
where $W, W_{1}, ..., W_{n}$ are smooth functions of $q^{a}$, $W_{,a}= \partial_{q^{a}}W$, $h$ is an arbitrary constant and $c=(c_{1}, ..., c_{n})$ are integration constants. Separable Hamiltonian systems form a large class of integrable systems and, moreover, the additive separation of the H-J equation is related to the multiplicative separation of the corresponding Helmholtz (or Schr\"{o}dinger) equation. The separation of variables in the H-J equation, corresponding to a natural Hamiltonian $H= \frac{1}{2}\gamma^{ab}(q)p_{a}p_{b} +V(q)$ with a kinetic metric of any signature, is intrinsically characterized by geometrical objects on the Riemannian configuration manifold, i.e. Killing vectors (KVs), Killing tensors (KTs), and Killing webs. The intrinsic characterization in terms of Riemannian geometry of the additive separation of variables in the H-J equation is discussed e.g. in \cite{Benenti 1997}, \cite{Tsiganov 2000}, \cite{Benenti 2002} and references cited therein. The H-J theory in the context of the moving frames formalism of E. Cartan is discussed in \cite{Adlam}.

One application of the H-J theory which is relevant to the present paper is the determination of the autonomous conservative dynamical systems with two degrees of freedom which
are superintegrable with one cubic FI (CFI) and either one linear FI (LFI) or a QFI. It is found in \cite{Marquette Winternitz 2008} that the case of LFIs gives the
well-known cases of the harmonic oscillator and the Kepler potential, while
the case of QFIs gives five irreducible potentials whose finite trajectories
are all closed. In another relevant work \cite{McLenaghan2004}, concerning the classification of autonomous CFIs
of autonomous Hamiltonians with two degrees of freedom, the authors classify
the non-trivial third order KTs using the group invariants
of KTs defined on pseudo-Riemannian spaces of constant curvature
under the action of the isometry group. Higher order FIs are also discussed in \cite{Tsiganov 2008}. In all cases mentioned above, the studies concern autonomous Hamiltonians and autonomous FIs (see e.g. \cite{Fokas 1979}, \cite{Fokas 1980}, \cite{Kalnins 1980}, \cite{Evans 1990}, \cite{McLenaghan 2002}).

\subsubsection*{The direct method}

The direct method applies to second order holonomic dynamical systems which are not necessarily conservative. In this method, one assumes a generic FI, say $I(t,q,\dot{q})$,
which is of a polynomial form in terms of the velocities $\dot{q}^{a}$ with unknown
coefficients and requires the condition $dI/dt=0.$ This condition leads to a
system of PDEs involving the unknown coefficients (tensors) of $I$ together with the
elements which characterize the dynamical system, that is, the potential $V$
and the non-conservative generalized forces $F^{a}$. The solution of this system is done as
in the H-J method, that is, in terms of the geometrical objects on the Riemannian
configuration manifold, i.e. the collineations (KVs,
Homothetic vectors (HVs), Affine vectors (AVs), Projective Collineations (PCs) ) and KTs of the appropriate order. It
appears that the direct method has been introduced for the first time by Bertrand \cite{Bertrand 1852} in the study of integrable surfaces and later used by Whittaker \cite{Whittaker} in the determination of the integrable autonomous
conservative Newtonian systems with two degrees of freedom. In the course of time,
this method has been used and extended by various authors (see e.g. \cite{Katzin 1973}, \cite{Katzin 1974}, \cite{Tsamparlis 2020B}, \cite{Katzin 1981}, \cite{Hall 1983}, \cite{Horwood 2007}, \cite{Tsamparlis 2020}).

Most studies consider the integrability of autonomous conservative
dynamical systems with two degrees of freedom. These systems admit already the
Hamiltonian QFI; therefore, in order to establish their
integrability one needs one more independent autonomous FI. A review of the known integrable and superintegrable autonomous conservative
dynamical systems with two degrees of freedom in terms of QFIs is given in \cite{Hietarinta 1987} and a more extended one including time-dependent QFIs in \cite{MitsTsam sym}. It is to be noted that these reviews do not contain all integrable/superintegrable  dynamical systems of this type, because the general solution of one of the equations
resulting from the condition $dI/dt=0$, the Bertrand-Darboux equation, is not known.

In the review paper \cite{Hietarinta 1987}, as a rule,  the integrability of
the considered dynamical systems is
established in terms of FIs which are autonomous and quadratic. The time-dependent FIs are totally absent, whereas there are occasional
references mainly to CFIs and to a lesser extent to quartic FIs (QUFIs). However, as it has been indicated above, the time-dependent FIs are equally appropriate for establishing integrability \cite{Kozlov 1983, Vozmishcheva 2005}. The same applies to a greater degree for the higher order FIs. These two types of FIs are not usually considered because their determination is difficult, especially, when algebraic methods are employed. However, this does not apply to the geometric method where one uses the general results of differential geometry concerning the collineations (symmetries) of the kinetic metric to compute the FIs. An early example in this direction determines the time-dependent FIs of higher order of the geodesic equations \cite{Katzin 1981}.

Concerning the solution of the system of PDEs resulting from the condition $dI/dt=0$, there are two methods. The algebraic method in which one solves the differential equations following the standard approach (see e.g. \cite{Ranada 1997}, \cite{Hietarinta 1987}, \cite{Daskaloyannis 2006}); and the  geometric
method in which one `solves' the system of PDEs in terms of the collineations of the metric defined by the kinetic energy (kinetic metric) of the dynamical system (see e.g. \cite{Katzin 1981}, \cite{Horwood 2007}, \cite{Tsamparlis 2020}). In the present work, we follow the geometric method.

In the present paper, the main new result is Theorem \ref{thm.mFIs} which generalizes the results of \cite{Katzin 1981} to the case of autonomous holonomic dynamical systems. It turns out that the geodesic equations is a special case in which the system of PDEs resulting from the condition $dI/dt=0$ is directly integrable. Using this Theorem we show how one takes known results concerning integrable autonomous conservative systems in a unified method. Furthermore, we present new integrable autonomous dynamical systems with two degrees of freedom which admit only a CFI. Finally, we show how a dynamical system which was considered to be integrable is in fact superintegrable because it admits an additional time-dependent FI.

\section{The conditions for an $m$th-order FI of an autonomous dynamical system}

\label{sec.higher.order.FIs}

We consider the autonomous holonomic dynamical system
\begin{equation}
\ddot{q}^{a}=-\Gamma _{bc}^{a}\dot{q}^{b}\dot{q}^{c}-Q^{a}(q)
\label{eq.hfi1}
\end{equation}%
where $\Gamma _{bc}^{a}$ are the coefficients of the Riemannian connection of
the kinetic metric $\gamma _{ab}(q)$ defined by the kinetic energy of the
system and $-Q^{a}(q)$ are the generalized forces.

We look for $m$th-order FIs of the form
\begin{equation}
I^{(m)}=\sum_{r=0}^{m}M_{i_{1}i_{2}...i_{r}}\dot{q}^{i_{1}}\dot{q}^{i_{2}}...%
\dot{q}^{i_{r}}=M+M_{i_{1}}\dot{q}^{i_{1}}+M_{i_{1}i_{2}}\dot{q}^{i_{1}}\dot{%
q}^{i_{2}}+...+M_{i_{1}i_{2}...i_{m}}\dot{q}^{i_{1}}\dot{q}^{i_{2}}...\dot{q}%
^{i_{m}}  \label{eq.hfi2}
\end{equation}%
where $M_{i_{1}...i_{r}}(t,q)$ with $r=0,1,...,m$ are totally symmetric $r$-rank tensors, the index $(m)$ denotes the order of the FI and the Einstein
summation convention is used.

The condition
\begin{equation}
\frac{dI}{dt}=0  \label{eq.hfi3}
\end{equation}%
leads to the following system of equations
\begin{eqnarray}
M_{(i_{1}i_{2}...i_{m};i_{m+1})} &=&0  \label{eq.hfi4a} \\
M_{i_{1}i_{2}...i_{m},t}+M_{(i_{1}i_{2}...i_{m-1};i_{m})} &=&0
\label{eq.hfi4b} \\
M_{i_{1}i_{2}...i_{r},t}+M_{(i_{1}i_{2}...i_{r-1};i_{r})}-(r+1)M_{i_{1}i_{2}...i_{r}i_{r+1}}Q^{i_{r+1}} &=&0,%
\enskip r=1,2,...,m-1  \label{eq.hfi4c} \\
M_{,t}-M_{i_{1}}Q^{i_{1}} &=&0  \label{eq.hfi4d}
\end{eqnarray}%
where round brackets indicate symmetrization of the enclosed indices, a comma indicates partial derivative and a semicolon Riemannian
covariant derivative. Equation (\ref{eq.hfi4a}) implies that $M_{i_{1}i_{2}...i_{m}}$ is an $m$th-order KT of the kinetic metric $\gamma_{ab}$.

Equations (\ref{eq.hfi4a}) - (\ref{eq.hfi4d})
must be supplemented with the integrability conditions $M_{,i_{1}t}=M_{,ti_{1}}$ and $M_{,[i_{1}i_{2}]}=0$ of the scalar function:
\begin{eqnarray}
M_{i_{1},tt}-2M_{i_{1}i_{2},t}Q^{i_{2}}+\left( M_{c}Q^{c}\right) _{,i_{1}}
&=&0  \label{eq.hfi5.1} \\
2\left( M_{[i_{1}|c|}Q^{c}\right) _{;i_{2}]}-M_{[i_{1};i_{2}],t} &=&0.
\label{eq.hfi5.2}
\end{eqnarray}
We note that square brackets indicate antisymmetrization of the enclosed indices; and indices enclosed between vertical lines are overlooked by symmetrization or antisymmetrization
symbols.

Equations (\ref{eq.hfi4a}) - (\ref{eq.hfi5.2}) constitute the system of
equations which has to be solved.

\section{Determination of the $m$th-order FIs}

\label{sec.direct.meth}

In order to solve the system of equations (\ref{eq.hfi4a}) - (\ref{eq.hfi5.2}%
) we assume a polynomial form in $t$ for both the $m$th-order KT $%
M_{i_{1}...i_{m}}(t,q)$ and the $r$-rank totally symmetric tensors $%
M_{i_{1}...i_{r}}(t,q)$, where $r=1,2,...,m-1$, with coefficients depending
only on $q^{a}$. That is, we assume that:\newline
a. The $\mathbf{m}${\textbf {th-order KT}} $M_{i_{1}...i_{m}}(t,q)$ has the form
\begin{equation}
M_{i_{1}...i_{m}}(t,q)=C_{(0)i_{1}...i_{m}}(q)+%
\sum_{N=1}^{n}C_{(N)i_{1}...i_{m}}(q)\frac{t^{N}}{N}  \label{eq.hfi5.3}
\end{equation}%
where $C_{(N)i_{1}...i_{m}}$, $N=0,1,...,n$, is a sequence of arbitrary $\mathbf{m}${\textbf{th-order KTs}} of the kinetic metric $\gamma _{ab}$ and $n$ is the
degree of the considered polynomial. \newline
b. The $\mathbf{r}${\textbf{-rank totally symmetric tensors}} (not in general KTs!) $%
M_{i_{1}...i_{r}}(t,q)$, where $r=1,2,...,m-1$, have the form
\begin{equation}
M_{i_{1}...i_{r}}(t,q)=%
\sum_{N_{r}=0}^{n_{r}}L_{(N_{r})i_{1}...i_{r}}(q)t^{N_{r}},\enskip r=1,2,...,m-1  \label{eq.hfi5.4}
\end{equation}%
where $L_{(N_{r})i_{1}...i_{r}}(q)$, $N_{r}=0,1,...,n_{r}$, are arbitrary %
$\mathbf{r}${\textbf{-rank totally symmetric tensors} }and $n_{r}$ is the
degree of the considered polynomial.

The degrees $n, n_{r}$ of the above polynomial expressions of $t$ may be
infinite.

Substituting (\ref{eq.hfi5.3}), (\ref{eq.hfi5.4}) in the system of equations
(\ref{eq.hfi4b}) - (\ref{eq.hfi5.2}) (equation (\ref{eq.hfi4a}) is
identically satisfied since $C_{(N)i_{1}...i_{m}}$ are assumed to be $m$th-order KTs) we find the
solution given in the following Theorem.

\section{The Theorem}

\label{sec.theorem}

\begin{theorem}
\label{thm.mFIs} The independent $m$th-order FIs of the dynamical system
(\ref{eq.hfi1}) are the following: \bigskip

\textbf{Integral 1.}
\begin{eqnarray*}
I_{n}^{(m)} &=&\left( -\frac{t^{n}}{n}L_{(n-1)(i_{1}...i_{m-1};i_{m})}-...-%
\frac{t^{2}}{2}%
L_{(1)(i_{1}...i_{m-1};i_{m})}-tL_{(0)(i_{1}...i_{m-1};i_{m})}+C_{(0)i_{1}...i_{m}}\right)
\dot{q}^{i_{1}}...\dot{q}^{i_{m}}+ \\
&&+\sum_{r=1}^{m-1}\left(
t^{n}L_{(n)i_{1}...i_{r}}+...+tL_{(1)i_{1}...i_{r}}+L_{(0)i_{1}...i_{r}}%
\right) \dot{q}^{i_{1}}...\dot{q}^{i_{r}}+s\frac{t^{n+1}}{n+1}+ \\
&&+L_{(n-1)c}Q^{c}\frac{t^{n}}{n}+...+L_{(1)c}Q^{c}\frac{t^{2}}{2}%
+L_{(0)c}Q^{c}t+G(q)
\end{eqnarray*}%
where $C_{(0)i_{1}...i_{m}}$, $L_{(N)(i_{1}...i_{m-1};i_{m})}$ for $%
N=0,1,...,n-1$ are $\mathbf{m}${\textbf{th-order KTs}}, $L_{(n)i_{1}...i_{m-1}}$ is an $\mathbf{(m-1)}${\textbf{th-order KT}}, $s$ is an arbitrary constant defined by the
condition
\begin{equation}
L_{(n)i_{1}}Q^{i_{1}}=s  \label{eq.FI1f}
\end{equation}%
while the vectors $L_{(N)i_{1}}$ and the {\textbf{totally symmetric tensors}} $%
L_{(A)i_{1}...i_{r}}$, $A=0,1,...,n$, $r=2,3,...,m-2$ satisfy the conditions
\begin{eqnarray}
L_{(n)(i_{1}...i_{m-2};i_{m-1})} &=&-\frac{m}{n}%
L_{(n-1)(i_{1}...i_{m-1};i_{m})}Q^{i_{m}}  \label{eq.FI1a} \\
L_{(k-1)(i_{1}...i_{m-2};i_{m-1})} &=&-\frac{m}{k-1}%
L_{(k-2)(i_{1}...i_{m-1};i_{m})}Q^{i_{m}}-kL_{(k)i_{1}...i_{m-1}},\enskip %
k=2,3,...,n  \label{eq.FI1b} \\
L_{(0)(i_{1}...i_{m-2};i_{m-1})}
&=&mC_{(0)i_{1}...i_{m-1}i_{m}}Q^{i_{m}}-L_{(1)i_{1}...i_{m-1}}
\label{eq.FI1c} \\
L_{(n)(i_{1}...i_{r-1};i_{r})}
&=&(r+1)L_{(n)i_{1}...i_{r}i_{r+1}}Q^{i_{r+1}},\enskip r=2,3,...,m-2
\label{eq.FI1d} \\
L_{(k-1)(i_{1}...i_{r-1};i_{r})}
&=&(r+1)L_{(k-1)i_{1}...i_{r}i_{r+1}}Q^{i_{r+1}}-kL_{(k)i_{1}...i_{r}},%
\enskip k=1,2,...,n,\enskip r=2,3,...,m-2  \label{eq.FI1e} \\
\left( L_{(n-1)c}Q^{c}\right) _{,i_{1}} &=&2nL_{(n)i_{1}i_{2}}Q^{i_{2}}
\label{eq.FI1g} \\
\left( L_{(k-2)c}Q^{c}\right) _{,i_{1}}
&=&2(k-1)L_{(k-1)i_{1}i_{2}}Q^{i_{2}}-k(k-1)L_{(k)i_{1}},\enskip k=2,3,...,n
\label{eq.FI1h} \\
G_{,i_{1}} &=&2L_{(0)i_{1}i_{2}}Q^{i_{2}}-L_{(1)i_{1}}.  \label{eq.FI1i}
\end{eqnarray}

\textbf{Integral 2.}
\begin{equation*}
I^{(m)}_{e}= \frac{e^{\lambda t}}{\lambda} \left(
-L_{(i_{1}...i_{m-1};i_{m})} \dot{q}^{i_{1}} ... \dot{q}^{i_{m}} + \lambda
\sum_{r=1}^{m-1} L_{i_{1}...i_{r}} \dot{q}^{i_{1}} ... \dot{q}^{i_{r}} +
L_{i_{1}}Q^{i_{1}} \right)
\end{equation*}
where $\lambda\neq0$, $L_{(i_{1}...i_{m-1};i_{m})}$ is an $m$th-order KT and
the remaining totally symmetric tensors satisfy the conditions
\begin{eqnarray}
L_{(i_{1}...i_{m-2};i_{m-1})}&=& -\frac{m}{\lambda}
L_{(i_{1}...i_{m-1};i_{m})} Q^{i_{m}} -\lambda L_{i_{1}...i_{m-1}}
\label{eq.FI2a} \\
L_{(i_{1}...i_{r-1};i_{r})}&=&(r+1) L_{i_{1}...i_{r}i_{r+1}} Q^{i_{r+1}}
-\lambda L_{i_{1}...i_{r}}, \enskip r=2,3,...,m-2  \label{eq.FI2b} \\
\left(L_{c}Q^{c}\right)_{,i_{1}}&=& 2\lambda L_{i_{1}i_{2}} Q^{i_{2}}
-\lambda^{2}L_{i_{1}}.  \label{eq.FI2c}
\end{eqnarray}
\end{theorem}

The above Theorem for $m=2$ reduces to Theorem 3 of \cite{Tsamparlis 2020B} for the QFIs of autonomous dynamical systems.

We note that the FI $I^{(m)}_{n}$ consists of two independent FIs of the same order $J^{(m,1)}_{\ell}$ and $J^{(m,2)}_{\ell}$ which for an even order $m=2\nu$ ($\nu\in\mathbb{N}$) are computed by the formulae ($\ell\in\mathbb{N}$):

a.
\begin{eqnarray}
J^{(m=2\nu,1)}_{\ell}&=& \left( -\frac{t^{2\ell}}{2\ell} L_{(2\ell-1)(i_{1}...i_{m-1};i_{m})} - ... - \frac{t^{2}}{2} L_{(1)(i_{1}...i_{m-1};i_{m})} +C_{(0)i_{1}...i_{m}} \right) \dot{q}^{i_{1}} ... \dot{q}^{i_{m}} + \notag \\
&& + \sum_{1\leq r \leq m-1}^{odd} \left( t^{2\ell-1} L_{(2\ell-1)i_{1}...i_{r}} + ... + t^{3}L_{(3)i_{1}...i_{r}} +tL_{(1)i_{1}...i_{r}} \right) \dot{q}^{i_{1}} ... \dot{q}^{i_{r}} + \notag \\
&& + \sum_{1\leq r \leq m-1}^{even} \left( t^{2\ell} L_{(2\ell)i_{1}...i_{r}} + ... + t^{2}L_{(2)i_{1}...i_{r}} +L_{(0)i_{1}...i_{r}} \right) \dot{q}^{i_{1}} ... \dot{q}^{i_{r}} + \notag \\
&& + \frac{t^{2\ell}}{2\ell}L_{(2\ell-1)c}Q^{c} + ... + \frac{t^{2}}{2}L_{(1)c}Q^{c} + G(q) \label{eq.FI4a}
\end{eqnarray}
where $C_{(0)i_{1}...i_{m}}$, $L_{(N)(i_{1}...i_{m-1};i_{m})}$ for $N=1,3,...,2\ell-1$ are $m$th-order KTs and the following conditions are satisfied
\begin{eqnarray}
L_{(2\ell)(i_{1}...i_{m-2};i_{m-1})}&=& -\frac{m}{2\ell} L_{(2\ell-1)(i_{1}...i_{m-1};i_{m})} Q^{i_{m}} \label{eq.FI4.1} \\
L_{(k-1)(i_{1}...i_{m-2};i_{m-1})}&=&-\frac{m}{k-1} L_{(k-2)(i_{1}...i_{m-1};i_{m})} Q^{i_{m}} -kL_{(k)i_{1}...i_{m-1}}, \enskip k=3,5,...,2\ell-1 \label{eq.FI4.2} \\
L_{(0)(i_{1}...i_{m-2};i_{m-1})}&=& mC_{(0)i_{1}...i_{m-1}i_{m}} Q^{i_{m}} -L_{(1)i_{1}...i_{m-1}} \label{eq.FI4.3} \\
L_{(2\ell)(i_{1}...i_{r-1};i_{r})}&=& (r+1) L_{(2\ell)i_{1}...i_{r}i_{r+1}} Q^{i_{r+1}}, \enskip r=3,5,...,m-3 \label{eq.FI4.4} \\ L_{(k-1)(i_{1}...i_{r-1};i_{r})}&=& (r+1) L_{(k-1)i_{1}...i_{r}i_{r+1}} Q^{i_{r+1}} -kL_{(k)i_{1}...i_{r}}, \enskip k=1,3,...,2\ell-1, r=3,5,...,m-3 \notag \\
\label{eq.FI4.5} \\
L_{(k-1)(i_{1}...i_{r-1};i_{r})}&=& (r+1) L_{(k-1)i_{1}...i_{r}i_{r+1}} Q^{i_{r+1}} -kL_{(k)i_{1}...i_{r}}, \enskip k=2,4,...,2\ell, \enskip r=2,4,...,m-2 \label{eq.FI4.6} \\
\left( L_{(2\ell-1)c}Q^{c} \right)_{,i_{1}} &=& 4\ell L_{(2\ell)i_{1}i_{2}}Q^{i_{2}} \label{eq.FI4.7} \\
\left( L_{(k-2)c}Q^{c} \right)_{,i_{1}} &=& 2(k-1)L_{(k-1)i_{1}i_{2}}Q^{i_{2}} -k(k-1)L_{(k)i_{1}}, \enskip k=3,5,...,2\ell-1 \label{eq.FI4.8} \\
G_{,i_{1}}&=& 2L_{(0)i_{1}i_{2}}Q^{i_{2}} -L_{(1)i_{1}}. \label{eq.FI4.9}
\end{eqnarray}

b.
\begin{eqnarray}
J^{(m=2\nu,2)}_{\ell}&=& \left( -\frac{t^{2\ell+1}}{2\ell+1} L_{(2\ell)(i_{1}...i_{m-1};i_{m})} - ... - \frac{t^{3}}{3} L_{(2)(i_{1}...i_{m-1};i_{m})} - tL_{(0)(i_{1}...i_{m-1};i_{m})} \right) \dot{q}^{i_{1}} ... \dot{q}^{i_{m}} + \notag \\
&& + \sum_{1\leq r \leq m-1}^{odd} \left( t^{2\ell} L_{(2\ell)i_{1}...i_{r}} + ... + t^{2}L_{(2)i_{1}...i_{r}} +L_{(0)i_{1}...i_{r}} \right) \dot{q}^{i_{1}} ... \dot{q}^{i_{r}}+ \notag \\
&& + \sum_{1\leq r \leq m-1}^{even} \left( t^{2\ell+1} L_{(2\ell+1)i_{1}...i_{r}} + ... + t^{3}L_{(3)i_{1}...i_{r}} +tL_{(1)i_{1}...i_{r}} \right) \dot{q}^{i_{1}} ... \dot{q}^{i_{r}} + \notag \\
&& + \frac{t^{2\ell+1}}{2\ell+1}L_{(2\ell)c}Q^{c} + ... + \frac{t^{3}}{3}L_{(2)c}Q^{c} + tL_{(0)c}Q^{c} \label{eq.FI4b}
\end{eqnarray}
where $L_{(N)(i_{1}...i_{m-1};i_{m})}$ for $N=0,2,...,2\ell$ are $m$th-order KTs and the following conditions are satisfied
\begin{eqnarray}
L_{(2\ell+1)(i_{1}...i_{m-2};i_{m-1})}&=& -\frac{m}{2\ell+1} L_{(2\ell)(i_{1}...i_{m-1};i_{m})} Q^{i_{m}} \label{eq.FI5.1} \\ L_{(k-1)(i_{1}...i_{m-2};i_{m-1})}&=& -\frac{m}{k-1} L_{(k-2)(i_{1}...i_{m-1};i_{m})} Q^{i_{m}} -kL_{(k)i_{1}...i_{m-1}}, \enskip k=2,4,...,2\ell \label{eq.FI5.2} \\
L_{(2\ell+1)(i_{1}...i_{r-1};i_{r})}&=& (r+1) L_{(2\ell+1)i_{1}...i_{r}i_{r+1}} Q^{i_{r+1}}, \enskip r=3,5,...,m-3 \label{eq.FI5.3} \\
L_{(k-1)(i_{1}...i_{r-1};i_{r})}&=& (r+1) L_{(k-1)i_{1}...i_{r}i_{r+1}} Q^{i_{r+1}} -kL_{(k)i_{1}...i_{r}}, \enskip k=1,3,...,2\ell+1, r=2,4,...,m-2 \notag \\
\label{eq.FI5.4} \\
L_{(k-1)(i_{1}...i_{r-1};i_{r})}&=& (r+1) L_{(k-1)i_{1}...i_{r}i_{r+1}} Q^{i_{r+1}} -kL_{(k)i_{1}...i_{r}}, \enskip k=2,4,...,2\ell, \enskip r=3,5...,m-3 \label{eq.FI5.5} \\
\left( L_{(2\ell)c}Q^{c} \right)_{,i_{1}} &=& 2(2\ell+1)L_{(2\ell+1)i_{1}i_{2}}Q^{i_{2}} \label{eq.FI5.6} \\
\left( L_{(k-2)c}Q^{c} \right)_{,i_{1}} &=& 2(k-1)L_{(k-1)i_{1}i_{2}}Q^{i_{2}} -k(k-1)L_{(k)i_{1}}, \enskip k=2,4,...,2\ell. \label{eq.FI5.7}
\end{eqnarray}

For $\nu=1 \implies m=2$ the QFIs (\ref{eq.FI4a}) and (\ref{eq.FI4b}) reduce to the QFIs $I_{(1)}$ and $I_{(2)}$ respectively found in Theorem 3 of \cite{Tsamparlis 2020B} (see also Appendix in \cite{Tsamparlis 2020B}).

Moreover, for an odd order $m=2\nu+1$ the independent FIs $J^{(m,1)}_{\ell}$, $J^{(m,2)}_{\ell}$ of the FI $I^{(m)}_{n}$ are given by the relations $J^{(2\nu+1,1)}_{\ell} = J^{(2\nu+2,1)}_{\ell}\left(M_{i_{1}...i_{m}}=0\right)$ and $J^{(2\nu+1,2)}_{\ell} = J^{(2\nu+2,2)}_{\ell}\left(M_{i_{1}...i_{m}}=0\right)$.

The $m$th-order FIs of geodesic equations follow from the application of Theorem \ref{thm.mFIs} to the case $Q^{a}=0$. It is found that in this case the integral $I_{e}^{(m)}=0$, whereas the Integral 1 is given by the expression (see eq. (2.10) in \cite{Katzin 1981})
\begin{equation}
I^{(m)}_{m} = \sum_{r=0}^{m} \sum_{b=0}^{r} \frac{(-t)^{r-b}}{(r-b)!} C_{(i_{1}...i_{b};i_{b+1}..i_{r})} \dot{q}^{i_{1}} \dot{q}^{i_{2}} ...\dot{q}^{i_{r}} = \sum_{r=0}^{m} \sum_{b=0}^{r} \frac{(-t)^{r-b}}{(r-b)!} C_{i_{1}...i_{b};i_{b+1}..i_{r}} \dot{q}^{i_{1}} \dot{q}^{i_{2}} ...\dot{q}^{i_{r}} \label{eq.hfi9}
\end{equation}
where the $b$-rank totally symmetric tensors $C_{i_{1}...i_{b}}$ satisfy the condition (see eq. (2.9) in \cite{Katzin 1981})
\begin{equation}
C_{(i_{1}...i_{b};i_{b+1}..i_{m+1})} =0, \enskip b=0,1,2...,m.
\label{eq.hfi7}
\end{equation}
We note that in the case of geodesic equations the totally symmetric tensors $C_{;i_{1}...i_{m}}$, $C_{(i_{1};i_{2}...i_{m})}$, $C_{(i_{1}i_{2};i_{3}...i_{m})}$, ..., $C_{i_{1}i_{2}...i_{m}}$ are $m$th-order KTs of $\gamma_{ab}$.

\section{Killing tensors (KTs) of spaces of constant curvature}

\label{sec.KTs.notes}

In a space of constant curvature that admits $n_{0}$ KVs (gradient and non-gradient) $X_{Ia}$ where $I=1,2,...,n_{0}$ all KTs of order $m$ are of the form (see e.g. \cite{Takeuchi 1983}, \cite{Thompson 1984}, \cite{Thompson 1986}, \cite{Eastwood}, \cite{Nikiting}, \cite{Horwood 2008})
\begin{equation}
K_{i_{1}...i_{m}}=\alpha
^{I_{1}...I_{m}}X_{I_{1}(i_{1}}X_{|I_{2}|i_{2}}...X_{|I_{m}|i_{m})}
\label{eq.syKT2}
\end{equation}%
where $\alpha^{I_{1}...I_{m}}$ are constants and $1\leq I_{1}\leq I_{2}\leq ...\leq I_{m}\leq n_{0}$. In (\ref{eq.syKT2}) in general the parameters $\alpha^{I_{1}...I_{m}}$ are not all independent.

In the following section, we apply this result in the case of the Euclidean plane $E^{2}$.

\section{The geometric quantities of $E^{2}$}

\label{sec.E2.geometry}

$E^{2}$ admits two gradient KVs $\partial_{x}, \partial
_{y}$ whose generating functions are $x, y$ respectively, and one
non-gradient KV (the rotation) $y\partial _{x}-x\partial _{y}$. These
vectors are written collectively as
\begin{equation}
L_{a}=\left(
\begin{array}{c}
b_{1}+b_{3}y \\
b_{2}-b_{3}x%
\end{array}%
\right)  \label{FL.15}
\end{equation}%
where $b_{1},b_{2},b_{3}$ are arbitrary constants, possibly zero.

\subsection{KTs of order 2 in $E^{2}$}

- The general KT of order 2 in $E^{2}$ is \cite{Adlam, Chanu 2006}
\begin{equation}
C_{ab}=\left(
\begin{array}{cc}
\gamma y^{2}+2\alpha y+A & -\gamma xy-\alpha x-\beta y+C \\
-\gamma xy-\alpha x-\beta y+C & \gamma x^{2}+2\beta x+B%
\end{array}%
\right)  \label{FL.14b}
\end{equation}
where $\alpha, \beta, \gamma, A, B, C$ are arbitrary constants.

- The vector $L_{a}$ generating KTs of $E^{2}$ of the form $%
C_{ab}=L_{(a;b)} $ is\footnote{%
Note that $L_{a}$ in (\ref{FL.14}) is the sum of the non-proper ACs of $%
E^{2} $ and not of its KVs which give $C_{ab}=0.$}
\begin{equation}
L_{a}=\left(
\begin{array}{c}
-2\beta y^{2}+2\alpha xy+Ax+(2C-a_{1})y+a_{2} \\
-2\alpha x^{2}+2\beta xy+a_{1}x+By+a_{3}%
\end{array}%
\right)  \label{FL.14}
\end{equation}
where $a_{1}, a_{2}, a_{3}$ are also arbitrary constants.

- The KTs $C_{ab}=L_{(a;b)}$ in $E^{2}$ generated from the vector (\ref%
{FL.14}) are
\begin{equation}
C_{ab}=L_{(a;b)}=\left(
\begin{array}{cc}
L_{x,x} & \frac{1}{2}(L_{x,y}+L_{y,x}) \\
\frac{1}{2}(L_{x,y}+L_{y,x}) & L_{y,y}%
\end{array}%
\right) =\left(
\begin{array}{cc}
2\alpha y+A & -\alpha x-\beta y+C \\
-\alpha x-\beta y+C & 2\beta x+B%
\end{array}%
\right).  \label{FL.14.1}
\end{equation}
Observe that these KTs are special cases of the general KTs (\ref{FL.14b})
for $\gamma =0$.

We note that the vector $L_{a}$ given by (\ref{FL.14}) depends on 8
parameters while the generated KT $L_{(a;b)}$ depends on five of them the $%
\alpha, \beta, A, B, C$. This is because the remaining $8-5=3$ parameters $%
a_{1}, a_{2}, a_{3}$ of the vector $L_{a}$ generate the KVs in $E^{2}$ which generate the zero KTs.

\subsection{KTs of order 3 in $E^{2}$}

- The general KT $C_{abc}$ of order 3 in $E^{2}$ has independent components \cite{McLenaghan2004, HorwoodMc}
\begin{eqnarray}
C_{111}&=& a_{1}y^{3} +3a_{2}y^{2} + 3a_{3}y +a_{4}  \notag \\
C_{112}&=& -a_{1}xy^{2} -2a_{2}xy +a_{5}y^{2} -a_{3}x +a_{8}y +a_{9}  \notag
\\
C_{221}&=& a_{1}x^{2}y +a_{2}x^{2} -2a_{5}xy -a_{8}x -a_{6}y +a_{10}
\label{eq.KT1} \\
C_{222}&=&-a_{1}x^{3} +3a_{5}x^{2} +3a_{6}x +a_{7}  \notag
\end{eqnarray}
where $a_{K}$ with $K=1,2,...,10$ are arbitrary constants.

- The reducible KT $C_{abc}= L_{(ab;c)}$ of order 3 in $E^{2}$ is generated by the
symmetric tensor
\begin{eqnarray}
L_{11}&=& 3b_{2}xy^{2} +3b_{5}y^{3} +3b_{3}xy +3(b_{10}+b_{8})y^{2} +b_{4}x
+3b_{15}y +b_{12}  \notag \\
L_{12}&=& -3b_{2}x^{2}y -3b_{5}xy^{2} -\frac{3}{2}b_{3}x^{2} -\frac{3}{2}%
(2b_{10}+b_{8})xy -\frac{3}{2}b_{6}y^2 +\frac{3}{2}(b_{9} -b_{15})x -\frac{3%
}{2}b_{11}y +b_{13}  \label{eq.KT2} \\
L_{22}&=& 3b_{2}x^{3} +3b_{5}x^{2}y +3b_{10}x^{2} +3b_{6}xy +3(b_{1}
+b_{11})x +b_{7}y +b_{14}  \notag
\end{eqnarray}
where $b_{1}, b_{2}, ..., b_{15}$ are arbitrary constants.

- The independent components of the generated KT are
\begin{eqnarray}
L_{(11;1)}&=& 3b_{2}y^{2} +3b_{3}y +b_{4}  \notag \\
L_{(11;2)}&=& -2b_{2}xy +b_{5}y^{2} -b_{3}x +b_{8}y +b_{9}  \notag \\
L_{(22;1)}&=& b_{2}x^{2}-2b_{5}xy -b_{8}x -b_{6}y +b_{1}  \label{eq.KT3} \\
L_{(22;2)}&=& 3b_{5}x^{2} +3b_{6}x +b_{7}.  \notag
\end{eqnarray}
We note that the KT (\ref{eq.KT3}) is just a subcase of the general KT (\ref%
{eq.KT1}) for $a_{1}=0$.

\section{Applications}

Theorem \ref{thm.mFIs} is covariant, independent of the dimension and
applies to a curved Riemannian space provided its geometric elements can be
determined. In that respect, it can be used to determine the higher order (time-dependent and autonomous) FIs of autonomous holonomic dynamical
systems. In the following, we demonstrate the application of the Theorem to
the rather simple case of Newtonian autonomous conservative dynamical systems with two degrees of
freedom which has been a research topic for many years. For these systems the kinetic metric $\gamma_{ab} =\delta_{ab}=diag(1,1)$ and $Q^{a}=-V^{,a}$ where $V(x,y)$ indicates the potential of the dynamical system.

As it has been mentioned, the known integrable and superintegrable systems of that type that admit QFIs are reviewed in \cite{Hietarinta 1987} and recently in \cite{MitsTsam sym}. Using the general Theorem \ref{thm.mFIs}, we shall show that: \newline
a. CFIs which have been determined by other methods follow as subcases directly from Theorem \ref{thm.mFIs}. \newline
b. New integrable potentials which admit only CFIs are found. \newline
c. Dynamical systems which were considered to be integrable admit an additional time-dependent CFI therefore are, in fact, superintegrable.

\subsection{Known CFIs}

In \cite{Fokas 1980}  the authors determined all potentials of the form $%
V=F(x^{2}+\nu y^{2})$, where $\nu $ is an arbitrary constant and $F$ an
arbitrary smooth function, that admit autonomous CFIs. They found the
following three potentials\footnote{%
There is a misprint in the FI (3.15b) of \cite{Fokas 1980} where the $p_{1}=%
\dot{x}$ in the last term must be $p_{2}=\dot{y}$.} (see eqs. (3.15a),
(3.15b), (3.19) of \cite{Fokas 1980})
\begin{equation}
V_{(1a)}=\frac{1}{2}x^{2}+\frac{9}{2}y^{2},\enskip V_{(1b)}=\frac{1}{2}x^{2}+\frac{1}{18}y^{2},\enskip V_{(1c)}=(x^{2}-y^{2})^{-2/3}.  \label{Fok1}
\end{equation}%
Using Theorem \ref{thm.mFIs} we found the new superintegrable\footnote{It is superintegrable because it is of the separable form $V(x,y)= F_{1}(x) +F_{2}(y)$, where $F_{1}, F_{2}$ are arbitrary smooth functions. It is well-known \cite{MitsTsam sym} that such potentials admit also the QFIs $I_{1}= \frac{1}{2}\dot{x}^{2} +F_{1}(x)$ and $I_{2}= \frac{1}{2}\dot{y}^{2} +F_{2}(y)$.} potential
\begin{equation}
V_{1}= c_{0}(x^{2} +9y^{2}) +c_{1}y \label{Fok2}
\end{equation}
where $c_{0}, c_{1}$ are arbitrary constants, which admits the associated CFI
\begin{equation}
J_{1}= (x\dot{y} -y\dot{x})\dot{x}^{2} -\frac{c_{1}}{18c_{0}} \dot{x}^{3} +\frac{c_{1}}{3}x^{2}\dot{x} +6c_{0}x^{2}y\dot{x} - \frac{2c_{0}}{3}x^{3}\dot{y} \label{Fok3}
\end{equation}
and the integrable potential
\begin{equation}
V_{2}= k(x^{2}-y^{2})^{-2/3} \label{Fok4}
\end{equation}
where $k$ is an arbitrary constant, which admits the CFI
\begin{equation}
J_{2}= \left( x\dot{y} -y\dot{x} \right) \left( \dot{y}^{2} -\dot{x}^{2} \right) +4V_{2}(y\dot{x} +x\dot{y}). \label{Fok6}
\end{equation}

We note that the potentials (\ref{Fok1}) are special cases of $V_{1}, V_{2}$ as follows:
\[
V_{(1a)}= V_{1}\left( c_{1}=0, c_{0}=\frac{1}{2} \right), \enskip V_{(1b)}= V_{1}\left( x \leftrightarrow y; c_{1}=0, c_{0}=\frac{1}{18} \right), \enskip V_{(1c)} =V_{2}(k=1).
\]

Working in the same manner one recovers all known potentials which are integrable or superintegrable and admit higher order FIs (see e.g. \cite{Tsiganov 2000}, \cite{Tsiganov 2008}, \cite{Karlovini 2000},\cite{ Karlovini 2002}).

\subsection{New integrable potentials}

Using Theorem \ref{thm.mFIs} we found the new integrable potential
\begin{equation}
V_{3}= \frac{k_{1}}{(a_{2}y-a_{5}x)^{2}} +\frac{k_{2}}{r} + \frac{k_{3}(a_{2}x+a_{5}y)}{r(a_{2}y-a_{5}x)^{2}} \label{new1}
\end{equation}
where $k_{1}, k_{2}, k_{3}, a_{2}, a_{5}$ are arbitrary constants and $r=\sqrt{x^{2}+y^{2}}$, which admits the CFI
\begin{eqnarray}
J_{3} &=&(x\dot{y}-y\dot{x})^{2}(a_{2}\dot{x}+a_{5}\dot{y}) +\frac{2k_{1}r^{2}%
}{(a_{2}y-a_{5}x)^{2}}(a_{2}\dot{x}+a_{5}\dot{y}) -\frac{k_{2}(a_{2}y-a_{5}x)%
}{r}(x\dot{y}-y\dot{x})+ \notag \\
&&+\frac{k_{3}r}{a_{2}y-a_{5}x}(a_{2}\dot{y}-a_{5}\dot{x}) -\frac{k_{3}(a_{2}x+a_{5}y)}{r(a_{2}y-a_{5}x)}(x\dot{y}-y\dot{x}) +\frac{2k_{3}(a_{2}x+a_{5}y)r}{(a_{2}y-a_{5}x)^{2}}(a_{2}\dot{x} +a_{5}\dot{y}). \label{new2}
\end{eqnarray}%

Furthermore, for $a_{5}=0$ this potential becomes the superintegrable potential (see last Table of \cite{MitsTsam sym})
\begin{equation}
V_{4}= V_{3}(a_{5}=0) =\frac{c_{1}}{y^{2}} +\frac{c_{2}}{r} +\frac{c_{3}x}{ry^{2}} \label{new3}
\end{equation}
where $c_{1}=\frac{k_{1}}{a_{2}^{2}}$, $c_{2}=k_{2}$, $c_{3}=\frac{k_{3}}{a_{2}}$ are arbitrary constants, which admits the CFI
\begin{equation}
J_{4}=J_{3}(a_{5}=0) = \left(x\dot{y} - y\dot{x} \right)^{2} \dot{x} -\frac{c_{2}y}{r}\left(x\dot{y} - y\dot{x} \right) +\frac{2c_{1}r^{2}}{y^{2}}\dot{x}+ \frac{c_{3} x(2x^{2} +3y^{2})}{ry^{2}}\dot{x} +\frac{c_{3}y}{r} \dot{y}. \label{new4}
\end{equation}
We note that under the transformation $c_{1}=B+C, c_{2}=A, c_{3}=C-B$, where $A, B, C$ are the new constants, the potential $V_{4}$ becomes
\begin{equation*}
V_{4}=\frac{A}{r} + \frac{B}{r(r+x)} +\frac{C}{r(r-x)} \label{new5}
\end{equation*}
which is the potential (3.2.36) of \cite{Hietarinta 1987}.

For $k_{2}=0$ we have the special potential
\begin{equation}
V_{5} =V_{3}(k_{2}=0)= \frac{k_{1}}{(a_{2}y-a_{5}x)^{2}} + \frac{k_{3}(a_{2}x+a_{5}y)}{r(a_{2}y-a_{5}x)^{2}} \label{eq.su1}
\end{equation}
which admits the additional time-dependent CFI
\begin{equation}
J_{5}= -tJ_{3}(k_{2}=0) +(a_{2}x +a_{5}y)(x\dot{y} -y\dot{x})^{2} + \frac{2k_{1}r^{2}(a_{2}x +a_{5}y)}{(a_{2}y -a_{5}x)^{2}} +\frac{2k_{3}r(a_{2}x +a_{5}y)^{2}}{(a_{2}y -a_{5}x)^{2}} +k_{3}r. \label{eq.su2}
\end{equation}
We conclude that $V_{5}$ is not just an integrable but a new superintegrable potential. This result illustrates the importance of the time-dependent FIs in the determination of the integrability/superintegrability.

\section{Conclusions}

Theorem  \ref{thm.mFIs} provides a general method for determining higher
order FIs of autonomous holonomic dynamical systems in a general space provided one
knows, or is able to calculate, the KTs of all orders --up to the order of the
FI-- of the kinetic metric. It is shown that an autonomous  dynamical system is possible to admit two families
of independent FIs of a given order. The results of
the Theorem are covariant and do not depend on the number of degrees of
freedom of the dynamical system.

We have considered an application of the Theorem in the rather simple --but widely studied-- case of autonomous
conservative dynamical systems with two degrees of freedom. It is shown
that one is possible to obtain the known integrable potentials which have
been computed using other methods but the direct method in a simple, direct, and concrete geometrical
approach. Finally, we have given a new integrable potential whose
integrability is established only by means of CFIs, and we have shown the importance of the time-dependent FIs in the determination of the integrability/superintegrability of a dynamical system by finding a new superintegrable potential.

\section*{Acknowledgement}

We would like to thank the referee for pointing out useful remarks which improved the quality of the paper.

\section*{Data Availability}

The data that supports the findings of this study are available within the article.

\section*{Conflict of interest}

The authors declare no conflict of interest.

\end{document}